\documentclass[draft]{agujournal2019}
\usepackage{url} 
\usepackage{lineno}
\usepackage[inline]{trackchanges} 
\usepackage{soul}
\usepackage{amsmath}
\usepackage{float}
\usepackage[justification=justified]{caption}
\usepackage{mathtools}
\usepackage{afterpage}

\linenumbers
\draftfalse

\journalname{Journal of Geophysical Research: Machine Learning and Computation}

\begin{document}
\nolinenumbers
%
%


\title{Simultaneous emulation and downscaling with physically-consistent deep learning-based regional ocean emulators} 

%
%




\authors{Leonard Lupin-Jimenez\affil{1}, Moein Darman\affil{1}, Subhashis Hazarika\affil{3}, Tianning Wu\affil{2}, Michael Gray\affil{2}, Ruyoing He\affil{2}, Anthony Wong\affil{3}, Ashesh Chattopadhyay\affil{1}}

\affiliation{1}{University of California, Santa Cruz, Department of Applied Mathematics}
\affiliation{2}{North Carolina State University, Marine, Earth, and Atmospheric Sciences}
\affiliation{3}{Fujitsu Research of America, Converging Technologies Laboratory}





\correspondingauthor{Ashesh Chattopadhyay}{aschatto@ucsc.edu}



\begin{keypoints}
\item An AI-based physically-consistent long-term regional emulator has been developed for the Gulf of Mexico region.
\item A deterministic and stochastic downscaling model has been developed to super-resolve the low resolution predictions to high-resolution. 
\item The performance of the emulation and downscaling has been compared with several baselines on short- and long-term metrics. 
\end{keypoints}

%
%

%
%


\begin{abstract}
  Building on top of the success in AI-based atmospheric emulation, we propose an AI-based ocean emulation and downscaling framework focusing on the high-resolution regional ocean over Gulf of Mexico. Regional ocean emulation presents unique challenges owing to the complex bathymetry and lateral boundary conditions as well as from fundamental biases in deep learning-based frameworks, such as instability and hallucinations. In this paper, we develop a deep learning-based framework to autoregressively integrate ocean-surface variables over the Gulf of Mexico at $8$ Km spatial resolution without unphysical drifts over decadal time scales and simulataneously downscale and bias-correct it to $4$ Km resolution using a physics-constrained generative model. The framework shows both short-term skills as well as accurate long-term statistics in terms of mean and variability.    

\end{abstract}

\section*{Plain Language Summary}

Data-driven models are promising tools for predicting ocean conditions and enhancing the details of these predictions. In this study, we applied advanced machine learning methods to model sea surface velocity and height in the Gulf of Mexico. To forecast broad ocean conditions, we used a method called Fourier Neural Operators (FNO), designed to balance computational efficiency with accuracy through a specialized loss function that combines grid and spectral space information. For creating high-resolution details from low-resolution data — a process called downscaling — we explored two different neural network architectures and compared their performance against simpler linear interpolation. This combination of forecasting and downscaling methods greatly improves the efficiency of ocean forecast and downscaling compared to numerical simulation with limited input variables. Our results highlight that these data-driven techniques can provide reliable, physics-aware predictions that can be useful for quick, localized analyses and in generating statistical predictions.

%
%

%


%
%
%

\section{Introduction}
The North Atlantic Ocean's western boundary current system (WBC), including the Loop Current (LC), Gulf Stream (GS), and Gulf Stream meander (GSM) play a significant role in controlling the Earth's ocean circulation, by transporting heat, salt, nutrients, and strongly influencing the global weather and climate system, including marine ecology. Modeling the regional ocean, e.g., in the Gulf of Mexico region (GoM), involves several challenges, starting with complex land boundaries, incorporation of lateral boundary conditions, which is usually computed from a global ocean model, and also eddy shedding events which is caused due to the interactions between the cool, subpolar circulation from the North and warm, subtropical circulation from the South. Accurately resolving the eddy shedding process in the GoM region has been a challenge for even high-resolution numerical ocean models in the past~\cite{dengo1993problem,chassignet2008gulf,ezer2016revisiting}. 

Recent years have seen widespread success in machine learning (ML-) based data-driven emulation of atmospheric dynamics~\cite{pathak2022fourcastnet, lam2022graphcast, bi2022pangu,guan2024lucie}, where the weather forecasting accuracy of such ML models have surpassed the accuracy of numerical weather prediction models, while being several thousand times faster. While several of these AI weather models such as FourCastNet~\cite{pathak2022fourcastnet}, GraphCast~\cite{lam2022graphcast}, and Pangu~\cite{bi2022pangu} eventually become unstable or unphysical, a few of the works around stability of these models for climate time scales have succcessfully demonstrated a long-term stable atmosphere with accurate climatology and variability~\cite{chattopadhyay2023long,watt2024ace2,guan2024lucie}. To scale such ML-based emulator approaches to the full Earth system, ocean emulators are essential. However, there have been limited work on building global or regional ocean emulators beyond investigating low-dimensional models predicting large-scale patterns at short time scales~\cite{wang2019medium,agarwal2021comparison}. Recently, we have demonstrated success in data-driven regional ocean modeling at very high resolution ($4$ Km) near the GoM and GS region where OceanNet~\cite{chattopadhyay2024oceannet} showed short-term prediction performance that was superior to numerical ocean models such as the Regional Ocean Modeling Systems (ROMS), while preserving long-term physical consistency. Similarly, Subel \textit{et al.}~\cite{subel2024building} and others~\cite{dheeshjith2024transfer,wang2024xihe,dheeshjith2024samudra} have demonstrated promising results in emulating the global ocean at different $CO_{2}$ forcings. 

In this paper, we build on the recent success in data-driven autoregressive ocean forecasting in OceanNet~\cite{chattopadhyay2024oceannet}, to extend it to multiple surface variables over climate time scales. Then, we investigate deep learning-based downscaling as a strategy to better resolve the GS emulated by the forecasting model. Downscaling and super resolution has been very popular in the weather and climate community~\cite{harris2022generative} where the focus has been primarily on increasing the fidelity of the forecasts from numerical models. Recently, Mardani \textit{et al.}~\cite{mardani2024residual} developed a generative model-based residual correction algorithm to downscale coarse-grained 25Km reanalysis to 2Km-scale fields (using observations) over Taiwan, wherein fine-scale convective structures were recovered. Such efforts in ocean dynamics have been largely absent. Here, we demonstrate both short- and long-term emulation of the surface ocean dynamics near the GoM region along with simultaneous downscaling at higher resolution. We highlight a few key important features in our framework. Unlike most work that downscale reanlaysis products into higher resolution observations, we autoregressively predict the surface ocean dynamics with an ML-based forecasting model and downscale the emulated fields to higher resolution. This is particularly difficult since autoregressive models have limited prediction skills, instability issues~\cite{chattopadhyay2023long, chattopadhyay2024oceannet}, and a tendency to become unphysical at long time scales. Furthermore, downscaling the predicted fields involve both super-resolution in the spatial fields as well as bias correction to account for error growth during autoregressive emulation. 

Our ML-based prediction and downscaling framework has the following features:

\begin{itemize}
    \item A long-term stable and physically consistent data-driven regional ocean emulator, i.e. a forecasting model (FC) trained on sea-surface height (SSH), sea-surface zonal and meridional velocities (SSU and SSV), and sea-surface kinetic energy (SSKE) from low-resolution (LR) GLORYS reanalysis data~\cite{GLORYS}. \\

    \item A deterministic and generative downscaling framework, DS model,  that super resolves and bias corrects the predicted fields from the FC model to high-resolution (HR) CNAPS reanalysis fields~\cite{CNAPS}.
\end{itemize}

In the rest of the paper, we describe the two datasets used for training the FC and DS models, the training and downscaling methodologies, principled structures used in the machine learning models to enforce physical consistency especially in the energy spectrum, and finally a results section that discusses the short- and long-term performance of the final downscaled fields with held-out CNAPS renalysis data.

\section{Datasets}
\begin{figure}[H]
    \centering
    \includegraphics[width=1\linewidth]{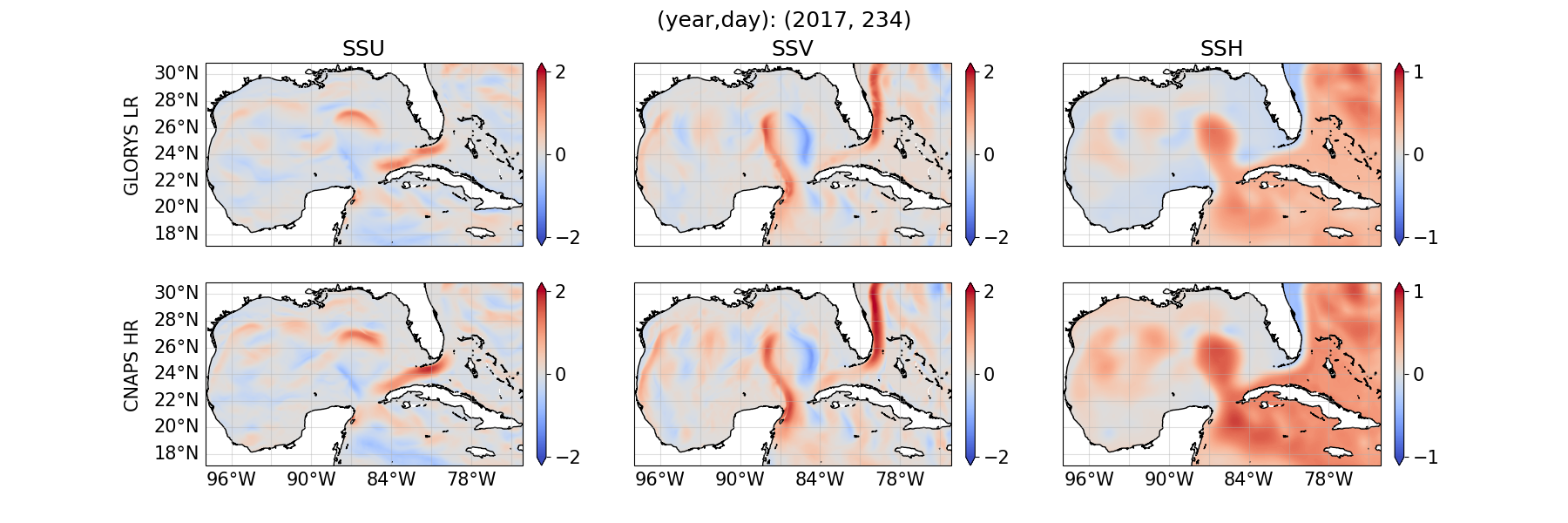}
    \caption{Example snaphosts for GLORYS low-resolution and CNAPS high-resolution datasets, for SSH, SSU, and SSV. There are differences between the fields, due to the differences in the reanalysis products.}
    \label{fig:field_example}
\end{figure}

\begin{figure}[H]
    \centering
    \includegraphics[width=1\linewidth]{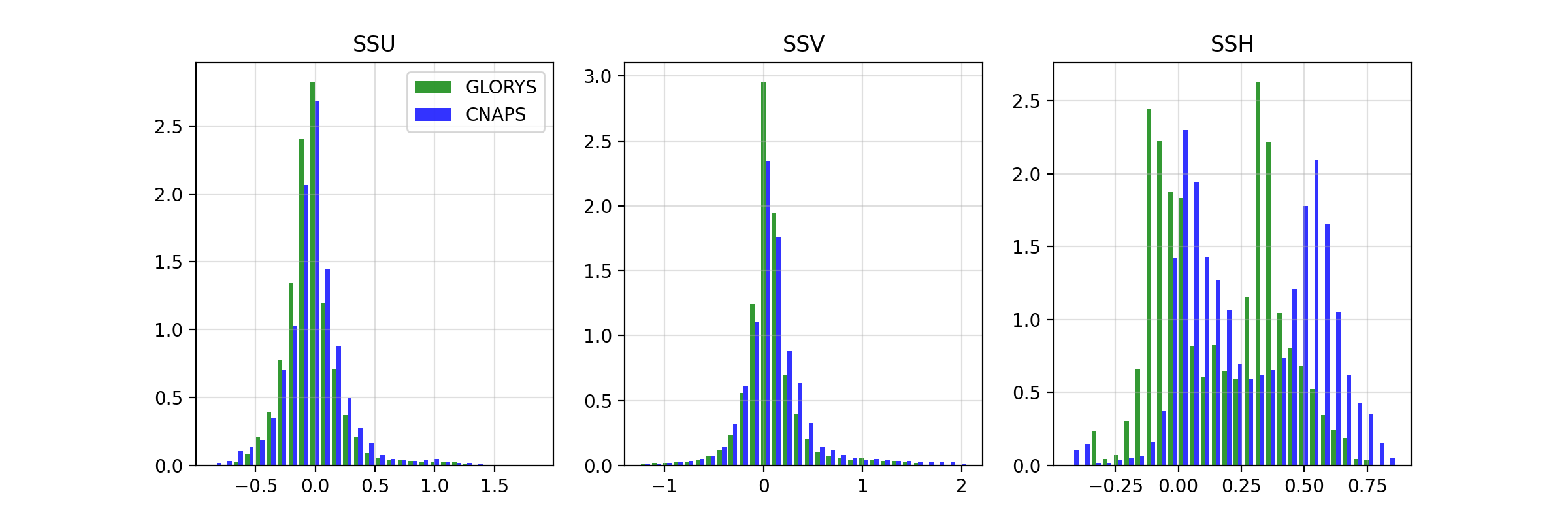}
    \caption{Distributions for GLORYS LR and CNAPS HR datasets, for SSH, SSU, and SSV.}
    \label{fig:field_example_hist}
\end{figure}

We used two reanalysis datasets in this paper. For training the autoregressive forecasting model, FC, we utilize LR data corresponding to SSH, SSU, SSV, and SSKE fields from the global ocean reanalysis, GLORYS~\cite{GLORYS}, at $8$ Km. We downscale the autoregressively emulated fields from FC to a HR $4$ Km regional reanalysis product, CNAPS~\cite{CNAPS}. Further details about the CNAPS reanalysis product can be obtained in ~\citeA{chattopadhyay2024oceannet}. In this paper, the autoregressive model integrates the surface ocean dynamics at a daily time scale. Unlike other atmospheric emulators, which are typically integrated at $6$ hourly temporal resolution, we integrated our ocean emulator daily, owing to the longer time scales of the oceanic processes. Figure~\ref{fig:field_example} shows the SSU, SSV, and SSH fields in GLORYS and CNAPS. While differences between the fields cannot be visually estimated, the distribution of the fields in Fig.~\ref{fig:field_example_hist} shows the difference, particularly in SSH. 



In our forecasting framework, we include a prognoastic variable (which is actually a diagnosed variable) that serves as an approximate measure of Sea Surface Kinetic Energy (SSKE). Although SSKE is not strictly conserved within our region of interest, incorporating it as a physical constraint enhances the model's accuracy by guiding predictions to better align with realistic ocean dynamics, especially with respect to the expected spectral properties of the region. Furthermore, using SSKE as a prognostic variable helps to maintain stable and energy-consistent outputs in the forecast.

\section{Methodology}
In the following sections, we outline the details of the FC and the DS model. Instead of forecasting at high-resolution by training on CNAPS data, as we had done in Chattopadhyay \textit{et al.}~\cite{chattopadhyay2024oceannet}, the FC model is trained on lower resolution GLORYS regional data. The choice of using a low-resolution forecasting model circumvents the cost of computational memory when using $4$ prognostic variables during training and also reduces spectral bias~\cite{chattopadhyay2023long}, thereby promoting stability of the FC model. Then, a physics-constrained generative (as well as a deterministic) downscaling model is trained to resolve higher resolution features in the predicted fields from the FC model to the CNAPS fields. The forecast and downscaling framework (FCDS) is shown in Figure \ref{fig:FCDS_pipeline} below. We have used a 2D Fourier neural operator (FNO) as the FC model and a modified variational autoencoder with a patch generative advserial network (PatchGAN-) based discriminator as well as a regular UNET-based architecture as the DS model. 

\begin{figure}[H]
    \centering
    \includegraphics[width=1.0\linewidth]{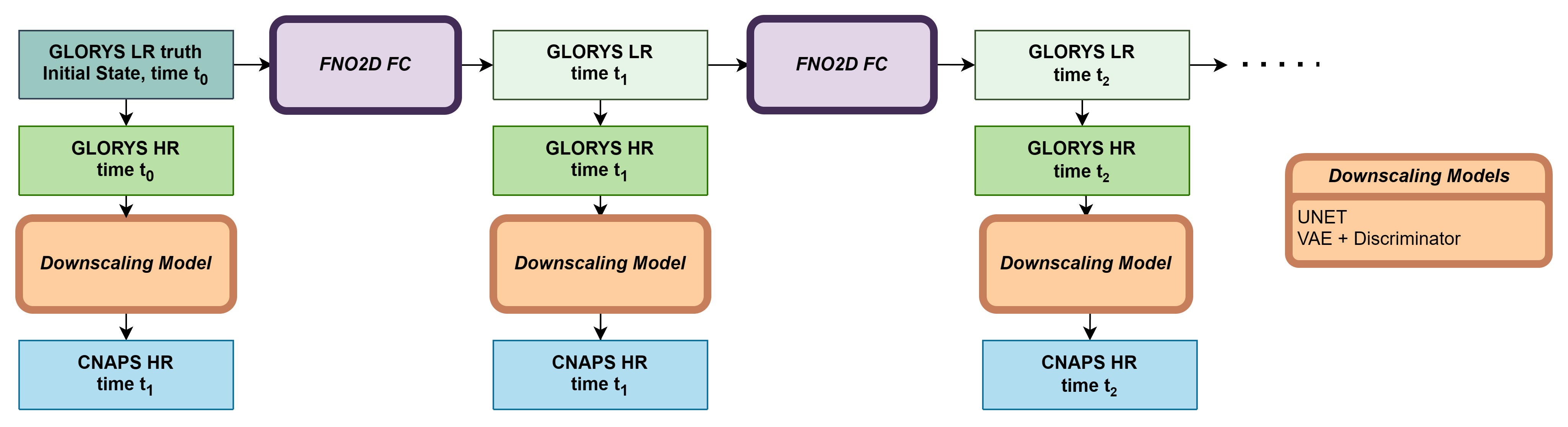}
    \caption{Framework of FC (left to right) and DS (top to bottom). Forecast inference is done from an initial low resolution GLORYS field state (GLORYS LR) at $t_0$. Before downscaling using our data driven models (top to bottom), the low resolution state is linearly interpolated to the high resolution CNAPS spatial grid.}
    \label{fig:FCDS_pipeline}
\end{figure}

\subsection{Loss functions}
For the training loss in the FC and DS models we use the weighted sum of grid and absolute spectral loss, shown below: 

\begin{equation}
    L_{\text{grid}} = \frac{1}{lmn} \sum_{c=1}^{l} \sum_{i=1}^{m}  \sum_{j=1}^{n} (y_{t;cij} - y_{p;cij})^2.
    \label{eq:grid_loss}
\end{equation}

\begin{equation}
    L_{\text{spectral, lon}} = \frac{1}{lm\hat{n}} \sum_{c=1}^{l} \sum_{i=1}^{m}  \sum_{\hat{j}=1}^{\hat{n}} (\hat{y}_{t;ci\hat{j}} - \hat{y}_{p;ci\hat{j}})^2.
    \label{eq:spectral_lon}
\end{equation}

\begin{equation}
    L_{\text{spectral, lat}} =  \frac{1}{l\hat{m}n} \sum_{c=1}^{l} \sum_{\hat{i}=1}^{\hat{m}}  \sum_{j=1}^{n} (\hat{y}_{t; c\hat{i}j} - \hat{y}_{p; c\hat{i}j})^2.
    \label{eq:spectral_lat}
\end{equation}

\begin{equation}
    L_{\text{total}} = (1-\lambda) L_{\text{grid}} + \lambda \frac{L_{\text{spectral, lon}} + L_{\text{spectral, lat}}}{2}.
    \label{eq:tot_loss}
\end{equation}


Here, $y_t$ and $y_p$ represent the ground truth and model predictions in grid space, respectively, while $\hat{y}_t$ and $\hat{y}_p$ represent the ground truth and model predictions in spectral space. In $y_{t;cij}$, $c$ is the channel, $i$ is the latitude index, and $j$ is the longitude index. $l$ denotes the number of output channels in the model. $m$ and $n$ refer to the number of latitude and longitude indexes in each axis of the grid space, while $\hat{m}$ and $\hat{n}$, respectively, indicate the corresponding number of latitude and longitude wave numbers in the spectral space. For grid points corresponding to land masses, the values of $y_t$ and $y_p$ are set to $0$, while for ocean, the values are set to $1$.

$L_\text{grid}$ represents the MSE loss in the original grid space of the fields. This is calculated for each prognostic variable represented in a channel $c$, and averaged at all points in the grid.  $L_{\text{spectral, lon}}$ and $L_{\text{spectral, lat}}$ are the spectral losses, with the Fourier transforms computed along the latitude in Eq.~(\ref{eq:spectral_lat}) and longitude in Eq.~(\ref{eq:spectral_lon}). Spectral loss allows us to reduce the spectral bias that leads to instabilities or unphysical drifts in autoregressive integration~\cite{chattopadhyay2023long}. In this work, since the Fourier transform is computed on a non-periodic region, we first mirror each field across the dimension we are computing the Fourier transform of, to enforce periodicity. We take the absolute value of the Fourier coefficients, $\hat{y}_t$ and $\hat{y}_p$ for the truth and the prediction, respectively.

$L_{\text{total}}$, the total loss, is computed as a weighted sum of the grid MSE and the spectral MSE loss, weighted by the parameter $\lambda$. This allows us to adjust the balance between the grid and spectral loss in the total loss computation. In practice, we found the optimal value of $\lambda$ for forecasting and downscaling is $0.2$. 

\subsection{FC model Training and Testing}
The proposed FC model is trained on GLORYS LR data, where SSH, SSU, SSV, and SSKE at day $t_i$ is used as the input and the same prognostic fields, at day, $t_{i+1}$ are used as labels. The SSKE is computed from SSU and SSV. The land-sea mask is also used as a constant input to the model. During training, the climatological lateral boundary conditions from the training data are enforced for the $4$ prognostic variables. The FC model in the FCDS frameworks is a two-dimensional Fourier neural operator (FNO2D) \cite{li2020fourier}. It has six Fourier layers, where $64$ Fourier modes are kept in each Fourier layer. The model is trained using the loss function given in Eq.~(\ref{eq:tot_loss}) using an ADAM optimizer. For training, we have utilized GLORYS fields from $1992$-$2018$. The autoregressive emulation skill of the FC model is validated with $50$ initial conditions, starting from $2019$ and into $2020$. 

\begin{figure}[ht]
    \centering
    \includegraphics[width=1\linewidth]{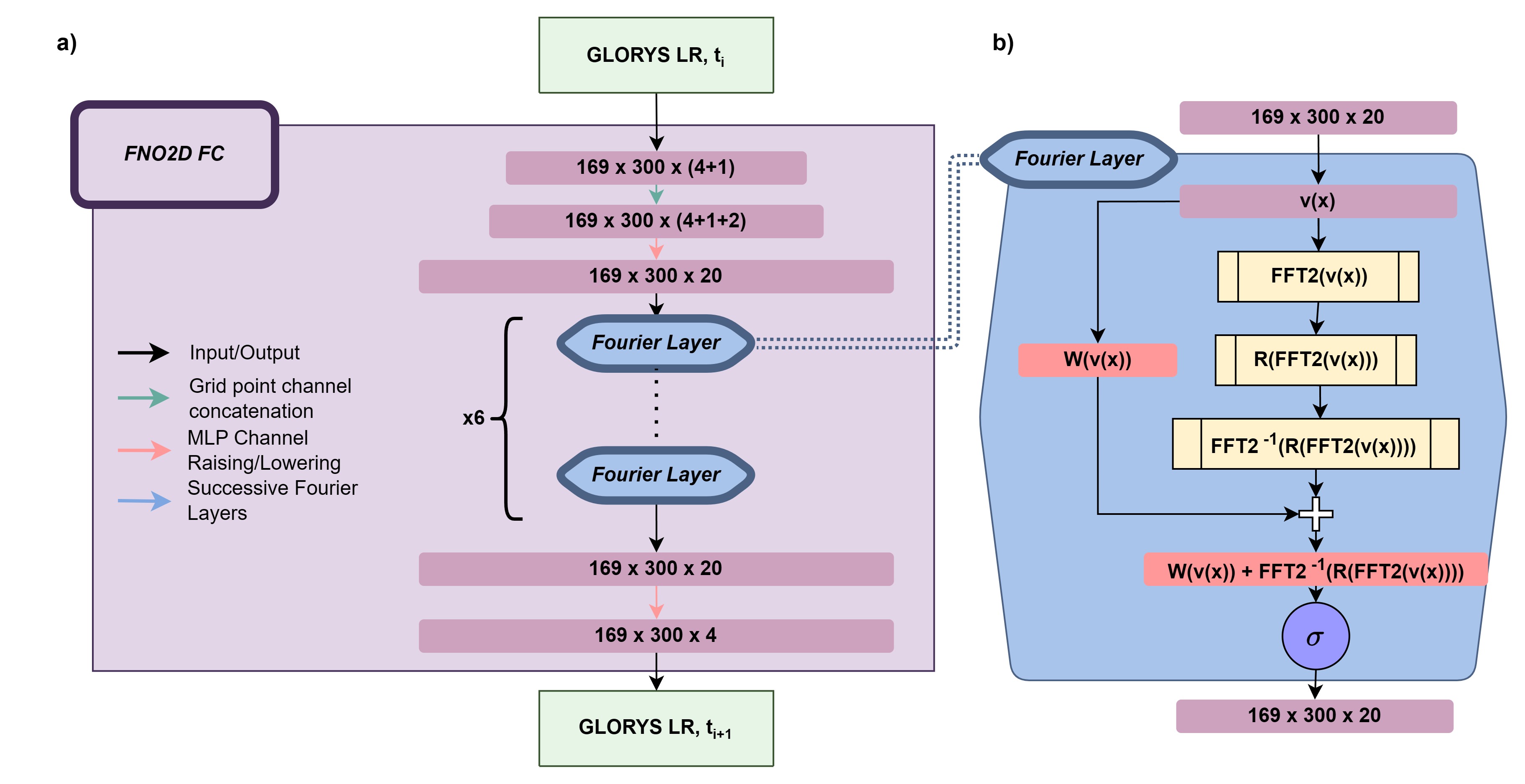}
    \caption{
    (a) A diagram of the full FNO2D network, with channel raising, Fourier layers, and channel lowering to the original dimensionality. Data is inputted with dimensionality (4+1), indicating the four channels and a single boolean 1/0 mask representing land or ocean. Then, the latitude and longitude coordinates are concatenated, giving (4+1+2). The channels are raised, and passed through 6 Fourier layers. Finally, the channels are lowered to the original 4-channel space, and the loss is computed from this output, show in equations Eqs.~(\ref{eq:grid_loss})-(\ref{eq:tot_loss}).(b) An individual Fourier layer. The input data passes through two separate channels: one, which performs a linear transformation $W(v(x))$ on the input, and another which performs a 2-D Fourier transform on the data $\text{FFT2}(v(x))$. In the second Fourier layer pipeline on the right of the diagram, the Fourier amplitudes are truncated, to remove higher wavenumber modes. A linear transform $R$ is then applied to this truncated form of the 2-D Fourier data, and then an inverse transform is applied. The linear transformation tensor is added to the Fourier operated tensor, and is passed through an activation function.}
    \label{fig:fno2d-detailed}
\end{figure}

\subsection{Downscaling Architectures}

We compare an UNET and a modified VAE architecture for downscaling in the DS model. The DS model super-resolves the low-resolution predicted variables from FC to high-resolution CNAPS fields. The DS model is trained offline to learn a map from the GLORYS prognostic fields to the CNAPS prognostic fields. For both the UNET and the VAE, the combined spatial grid and spectral loss function shown in Eq.~(\ref{eq:grid_loss})-Eq.~(\ref{eq:tot_loss}) is used to reconstruct the high-resolution features of the flow. 

\subsubsection{UNET}
\begin{figure}[H]
    \centering
    \includegraphics[width=1\linewidth]{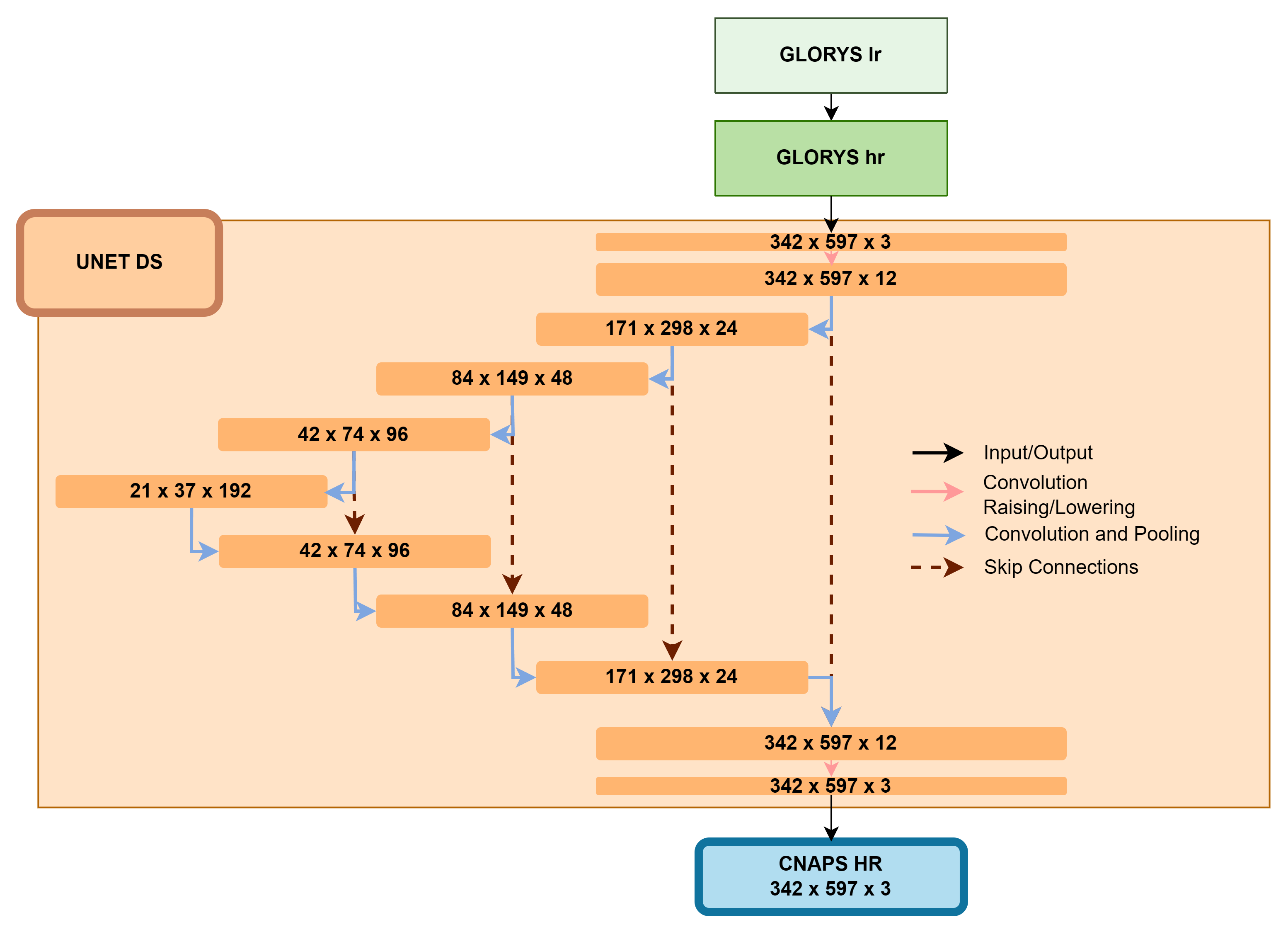}
    \caption{Diagram of UNET architecture used; successive layers shown top to bottom. The $x$ and $y$ sizes of each layer, as well as the number of channels $c$, are shown in the form in the form $x \times y \times c$. There are for contraction, which perform $3 \times 3$ kernel convolutions with ReLU activation, along with a max-pooling layer. Then, 4 layers of expansion are done, with skip connection concatenation, up-convolution $2\times2$ kernel convolution.}
    \label{fig:UNET_arch}
\end{figure}

The UNET architecture, illustrated in Fig. \ref{fig:UNET_arch}, is originally designed for image segmentation, but it has also shown significant promise in various other tasks, including image classification, regression, and downscaling. 

The network is structured in a U-shape, with an encoder-decoder configuration. It begins with a channel-raising convolutional layer, followed by four down-sampling layers consisting of convolutional operations and pooling layers, which progressively reduce the spatial resolution while increasing the number of feature channels. This encoder portion captures the high-level features of the input data by compressing spatial information into a compact latent representation.

The decoder portion of the network consists of four up-sampling layers with transposed convolutions, which progressively reconstruct the spatial resolution of the data. These layers use learned filters to expand the feature maps, enabling the network to reconstruct the output image with high spatial fidelity. The key feature of the UNET architecture is the use of skip connections between the corresponding layers of the encoder and decoder. These skip connections allow the model to retain fine-grained spatial information that might otherwise be lost during down-sampling. By directly passing feature maps from the encoder to the decoder, these connections help refine the output and prevent the loss of important structural details.

In our specific application, the UNET framework leverages its latent space representation and skip connections to refine low-resolution input data. By utilizing the detailed feature maps passed through the skip connections, the network can correct discrepancies and adjust for the differences between the low-resolution and high-resolution data. This ensures that the model is better equipped to map the low-resolution input to a high-resolution output, aligning the predictions more closely with the characteristics of the high-resolution data. This approach enhances the accuracy and detail of the downscaled output, making it more suitable for applications requiring fine spatial resolution.

\subsubsection{Variational Autoencoder with Adversarial Training}

The Variational Autoencoder (VAE) is designed to reconstruct high-resolution CNAPS data from low-resolution GLORYS data while learning a meaningful latent representation. The architecture consists of an encoder, a decoder, and a PatchGAN discriminator, trained in an adversarial framework.

The encoder processes the input through an initial convolutional layer, followed by several downsampling layers (\textit{DownBlocks}), which reduce spatial dimensions while increasing feature channels. The bottleneck layers (\textit{MidBlocks}) refine the feature maps before generating the latent distribution, represented by the mean $\mu$ and log variance $\log \sigma^2$. The latent sample $z$ is computed using the reparameterization trick:
\begin{equation}
z = \mu + \sigma \cdot \epsilon, \quad \sigma = \exp(0.5 \cdot \log \sigma^2), \quad \epsilon \sim \mathcal{N}(0, 1),
\label{eq:vae_reparam}
\end{equation}
where $\mathcal{N}(0,1)$ denotes the standard normal distribution. Sampling the latent variable $z$ using equation Eq.~(\ref{eq:vae_reparam}) ensures differentiability while mapping inputs to latent space, thus improving latent representation for reconstruction. Encoded feature maps are preserved as skip connections to enhance the decoder's reconstruction ability.

The decoder concatenates the latent sample with the skip-connected features and upsamples the spatial dimensions through \textit{UpBlocks}, mirroring the encoder's downsampling operations. Additional \textit{MidBlocks} further refine the reconstructed feature maps. A final convolutional layer produces the high-resolution output, cropped to remove padding applied during the encoder’s input processing.

The discriminator is a PatchGAN, designed to evaluate the realism of reconstructed images at the patch level. It processes inputs through sequential convolutional layers, progressively reducing spatial dimensions while predicting a grid of values. Each cell in the grid corresponds to a patch of the input, with higher values indicating greater realism. The architecture employs LeakyReLU activation and batch normalization, with the final layer producing the grid of predictions.

The training combines multiple loss functions to optimize the VAE and discriminator, represented by $\mathcal{V}$ and $\mathcal{D}$, respectively. The reconstruction loss incorporates spatial and spectral components and is defined as:
\begin{equation}
L_{\text{recon}} = L_{\text{total}} = (1-\lambda) L_{\text{grid}} + \lambda \frac{L_{\text{spectral, lon}} + L_{\text{spectral, lat}}}{2}.
\end{equation}

Latent space regularization is achieved via the Kullback-Leibler (KL) divergence:
\begin{equation}
L_{\text{KL}} = -\frac{1}{2} \sum \left( 1 + \log \sigma^2 - \mu^2 - \sigma^2 \right).
\end{equation}

The adversarial loss encourages the generator to produce realistic outputs classified as real by the discriminator:
\begin{equation}
L_{\text{adv}} = \frac{1}{N} \sum \left( \mathcal{D}(\mathcal{V}(X(t))) - 1 \right)^2,
\end{equation}
where $\mathcal{D}(\circ)$ represents the discriminator's output, with values in $(0,1)$; 0 indicates generated samples, and 1 indicates real samples.

The total generator loss \(L_{\text{gen}}\) combines the reconstruction loss \(L_{\text{recon}}\), the Kullback-Leibler divergence \(L_{\text{KL}}\) weighted by \(\beta_{\text{KL}}\), and the adversarial loss \(L_{\text{adv}}\) weighted by \(\lambda_{\text{adv}}\), where \(\beta_{\text{KL}}\) and \(\lambda_{\text{adv}}\) control the relative contributions of regularization and adversarial terms.

\begin{equation}
L_{\text{gen}} = L_{\text{recon}} + \beta_{\text{KL}} \cdot L_{\text{KL}} + \lambda_{\text{adv}} \cdot L_{\text{adv}}.
\end{equation}

The discriminator minimizes the following objective:
\begin{equation}
L_{\text{disc}} = \frac{1}{2} \left( L_{\text{real}} + L_{\mathcal{V}} \right),
\end{equation}
where:
\begin{align}
L_{\text{real}} &= \frac{1}{N_{\text{real}}} \sum_{i=1}^{N_{\text{real}}} \left( \mathcal{D}(X_{\text{hr},i}(t)) - 1 \right)^2, \\
L_{\mathcal{V}} &= \frac{1}{N_{\text{lr}}} \sum_{i=1}^{N_{\text{lr}}} \left( \mathcal{D}(\mathcal{V}(X_{\text{lr},i}(t))) - 0 \right)^2.
\end{align}
Here, $X_{\text{hr},i}(t)$ denotes the high-resolution CNAPS ground truth, and $\mathcal{V}(X_{\text{lr},i}(t))$ represents low-resolution VAE-reconstructed data.

Training alternates between optimizing the generator and the discriminator. The generator minimizes \(L_{\text{gen}}\) via gradient accumulation, while the discriminator minimizes \(L_{\text{disc}}\). The dynamic adjustment of the learning rates for both networks ensures stability, preventing either model from overpowering the other. This dynamic adjustment ensures convergence and produces high-quality realistic reconstructions.

\begin{figure}[H]
    \centering
    \includegraphics[width=1\linewidth]{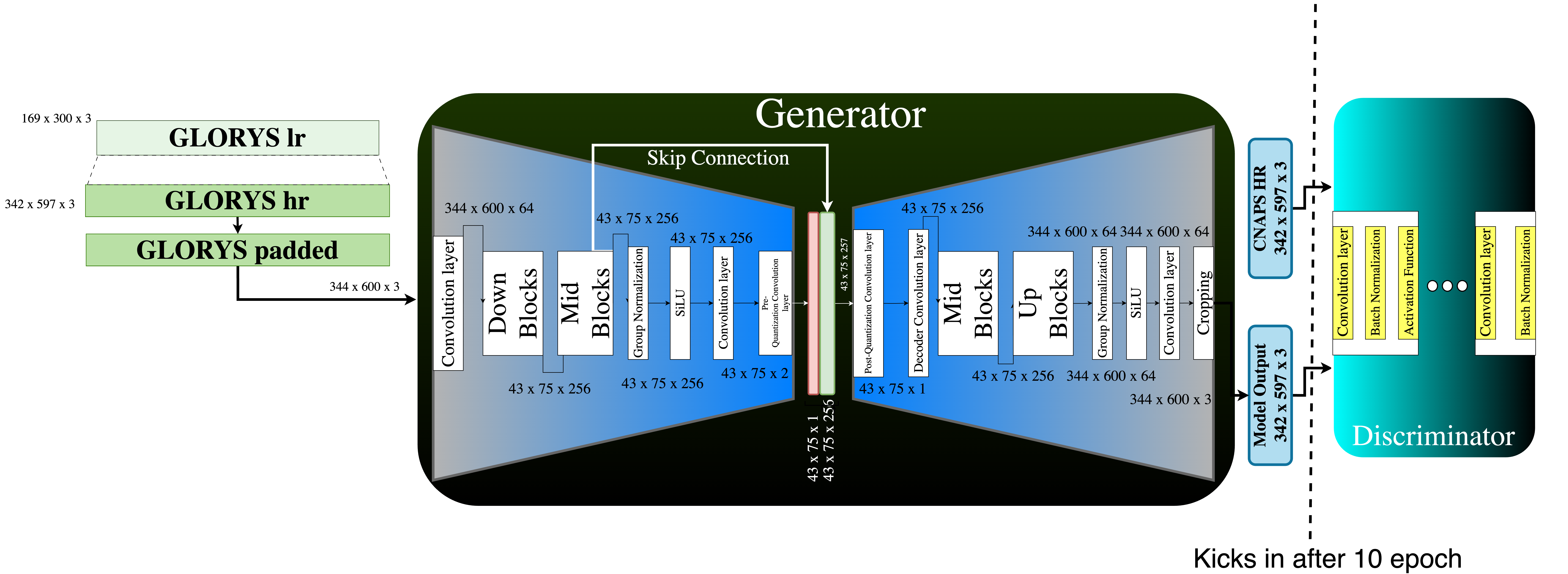}
    \caption{Diagram of the VAE architecture, illustrating the transformation from low-resolution GLORYS data input to high-resolution CNAPS output. The encoder maps the input to a latent space characterized by the mean $(\mu)$ and log variance $(\log \sigma^2)$, with the reparameterization trick enabling differentiable sampling. Skip connections retain critical spatial features to aid the decoder, which upsamples and reconstructs the high-resolution outputs. The PatchGAN discriminator evaluates the realism of reconstructions at the patch level. Training is guided by loss function comprising reconstruction, KL divergence, and adversarial components, producing high-quality realistic reconstructions, and robust latent space learning.}
    \label{fig:VAE}
\end{figure}

\subsubsection{Online fine tuning of the DS model for bias correction}
Unlike standard downscaling or super-resolution tasks, in this paper, we are downscaling the fields predicted by the autoregressive emulator which in itself has a model error and after a few integration time steps diverges from the true LR GLORYS fields, due to a combined effect of model error and sensitivity to initial conditions. In order to mitigate the bias between the diverging emulation and the high-resolution CNAPS fields, we fine-tune both the DS models using the same loss functions on training data from $2018$. The online fine tuning process evolves the autoregressive model from $50$ initial conditions in $2018$ for $30$ days. The offline-trained DS models are then fine-tuned to correct the bias between the emulation and CNAPS. There are several advantages in performing online fine tuning. One of the major ones is to account for and correc the discrepncy between the different meso-scale ocean parameterizations used in the physical model that was used to generate the reanalysis data in GLORYS and CNAPS. 

\section{Results}
In this section, we demonstrate the performance of the FCDS framework, using both UNET and VAE, with several different skill metrics for both short-term performance and long-term statistics. 

\subsection{Short-term skills of the FCDS framework}
\begin{figure}[H]
    \centering
    \includegraphics[width=1\linewidth]{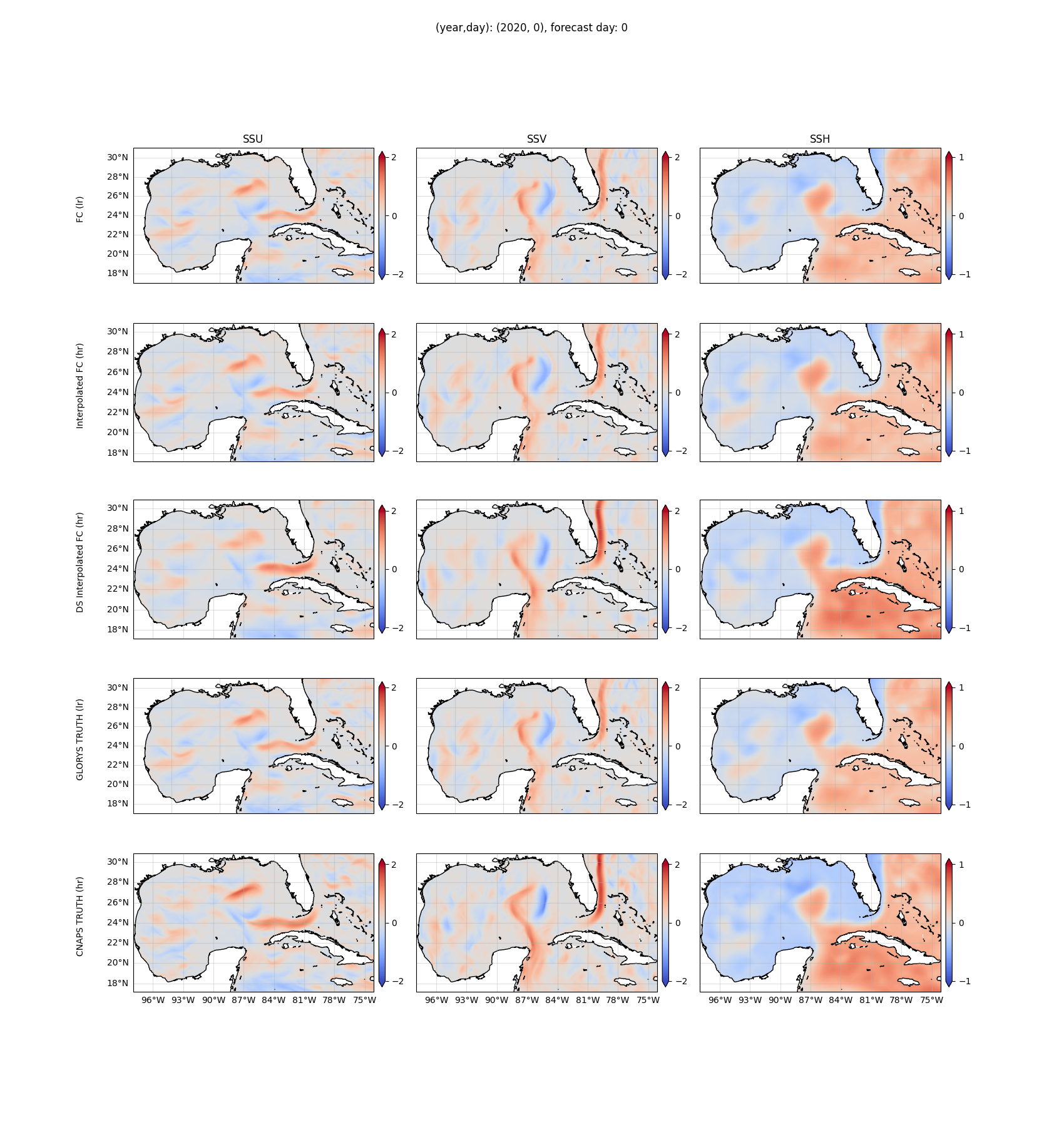}
    \caption{An output snapshot from the FCDS framework starting from a single initial condition. The first 3 rows represent the forecast and downscaling results. Rows starting from top to bottom are 1) the FNO2D low resolution forecast, 2) the high resolution interpolation, 3) the online-finetuned UNET model's downscaled forecast, 4) the low resolution interpolated GLORYS's true snapshot and 5) the high resolution CNAPS's  true snapshot. The columns are the SSH, SSU, and SSV fields.}
    \label{fig:fcds_day-0}
\end{figure}

\begin{figure}[H]
    \centering
    \includegraphics[width=1\linewidth]{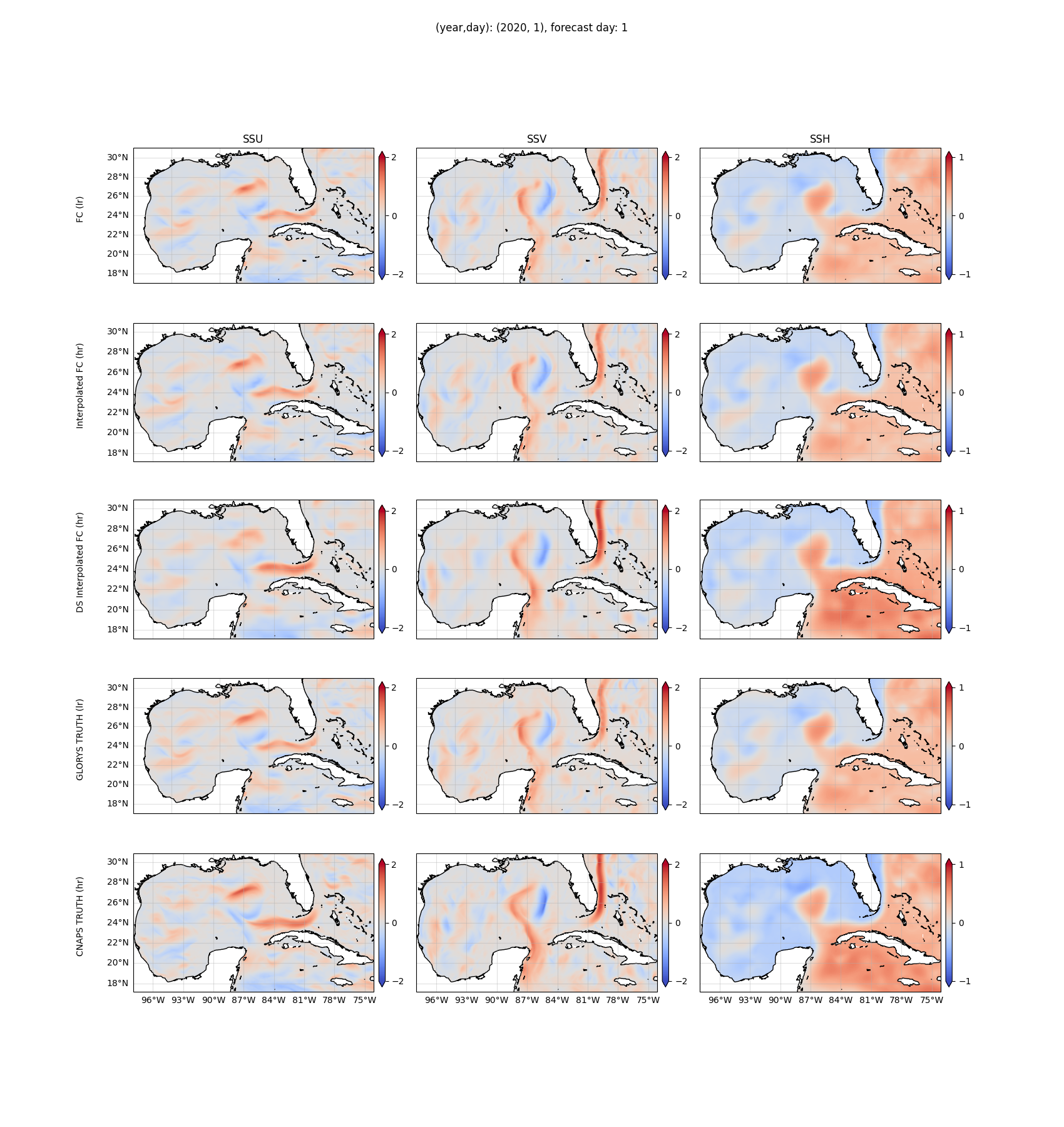}
    \caption{$1$ day prediction with the FCDS framework.}
    \label{fig:fcds_day-1}
\end{figure}

\begin{figure}[H]
    \centering
    \includegraphics[width=1\linewidth]{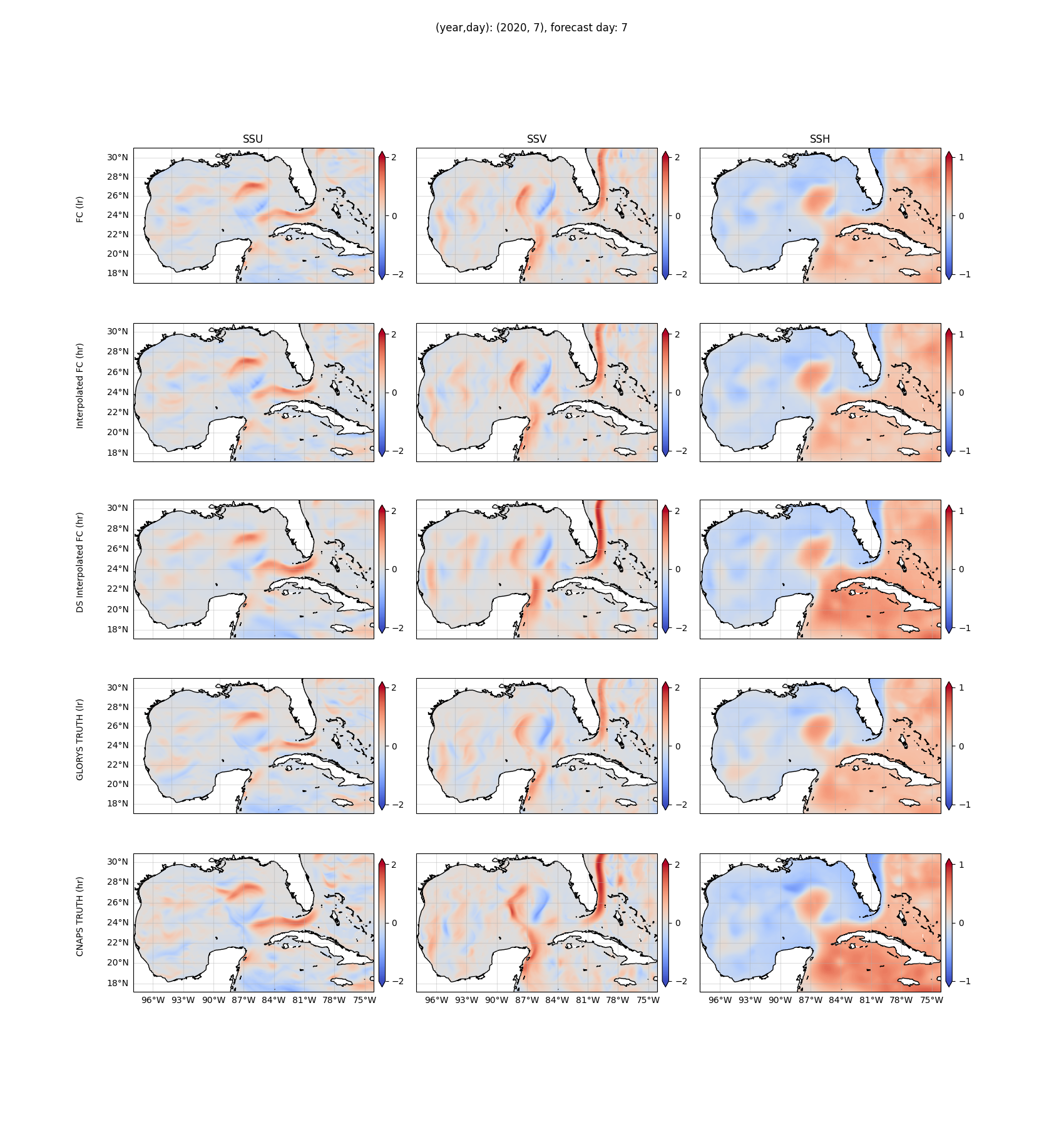}
    \caption{$7$ day prediction with the FCDS framework.}
    \label{fig:fcds_day-7}
\end{figure}

\begin{figure}[H]
    \centering
    \includegraphics[width=1\linewidth]{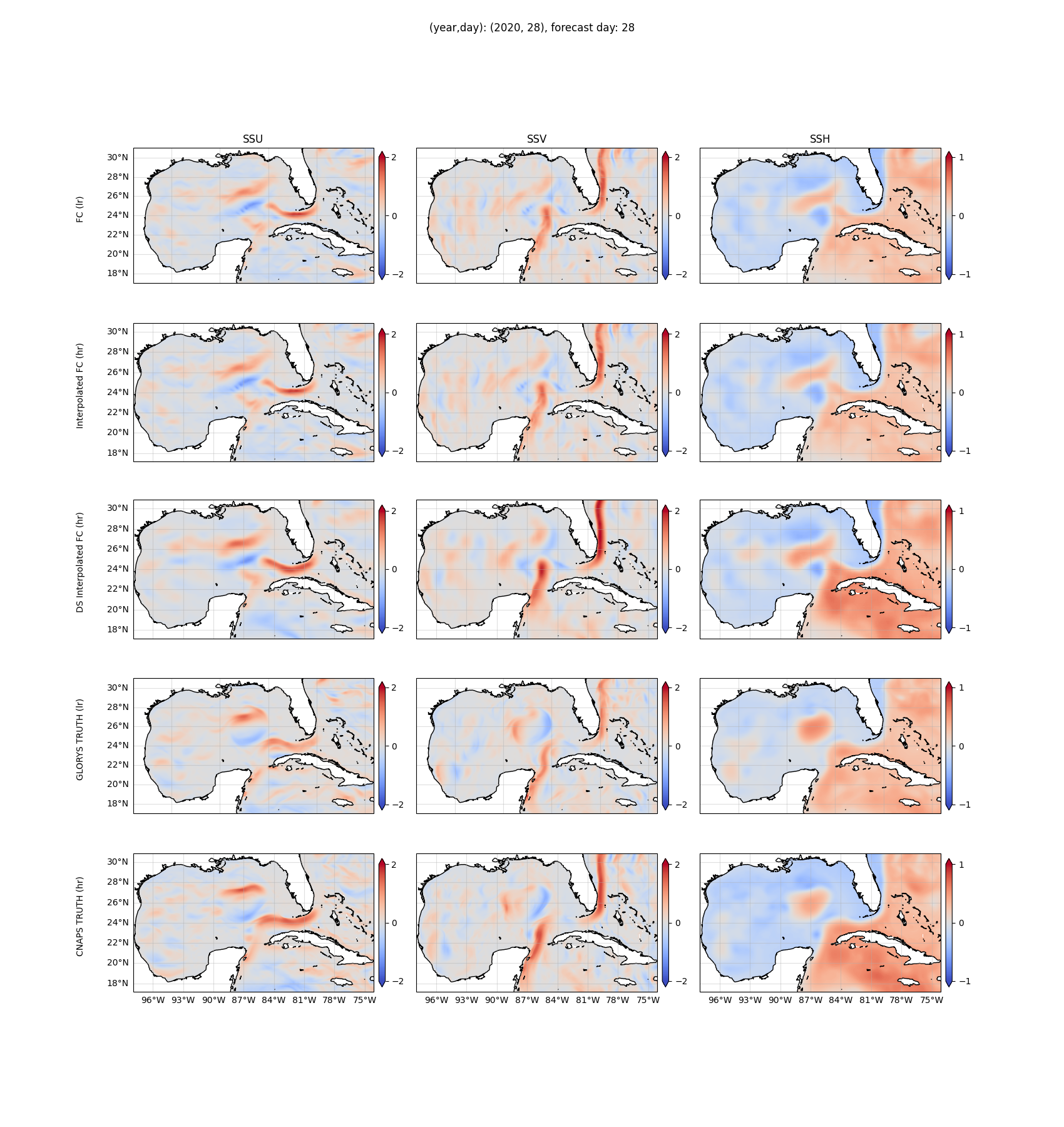}
    \caption{$28$ day prediction with the FCDS framework.}
    \label{fig:fig:fcds_day-28}
\end{figure}

\begin{figure}[H]
    \centering
    \includegraphics[width=1\linewidth]{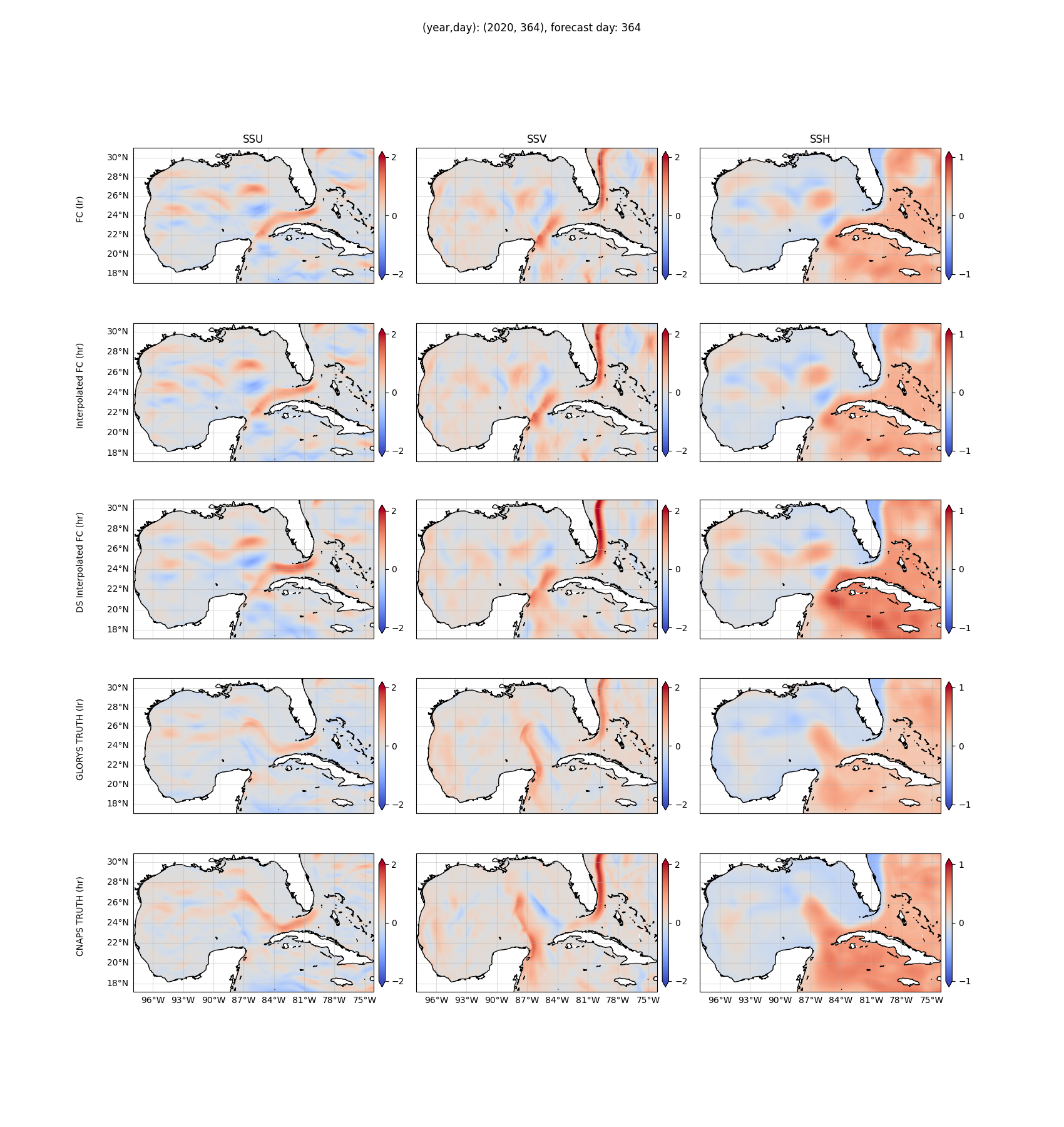}
    \caption{$365$ day prediction with the FCDS framework.}
    \label{fig:fcds_day-364}
\end{figure}

Here, we show that the FCDS emulated and downscaled fields for SSH, SSU, and SSV visually in Fig.~\ref{fig:fcds_day-0} to Fig.~\ref{fig:fcds_day-364}. Here, Fig.~\ref{fig:fcds_day-0} is the initial condition while Fig.~\ref{fig:fcds_day-364} shows the emulated and downscaled fields after a year. It must be noted that we do not expect the FCDS outputs to match the true high-resolution CNAPS fields after a year. The fields, however, do not go unstable and remain physically consistent. Each of the panels in Fig.~\ref{fig:fcds_day-0} to Fig.~\ref{fig:fcds_day-364} represent the emulated fields in low-resolution space using the FNO-based autoregressive model, a naive baseline super-resolution onto the CNAPS grid using bilininear interpolation, FCDS output, and the true GLORYS and CNAPS fields, from top to bottom. The forecast accuracy of the autoregressive model can be qualitatively assessed by comparing the first and the fourth panels, while the accuracy of the downscaled fields can be assessed by comparing the third and the fifth panel. Here, we show the snapshots of the fields starting from one initial condition. However, to compare skills, we have conducted emulation and downscaling over multiple initial conditions. In Fig.~\ref{fig:fcds_day-0} to Fig.~\ref{fig:fcds_day-364} we have shown the results with the online fine-tuned UNET-based DS model, which leads to the most performant FCDS framework. It is clear from the figures that the intensity of the GS is captured more accurately in the SSH field of the FCDS framework's output as compared to the baseline interpolation method. 

\begin{figure}[H]
    \centering
    \includegraphics[width=1.0\linewidth]{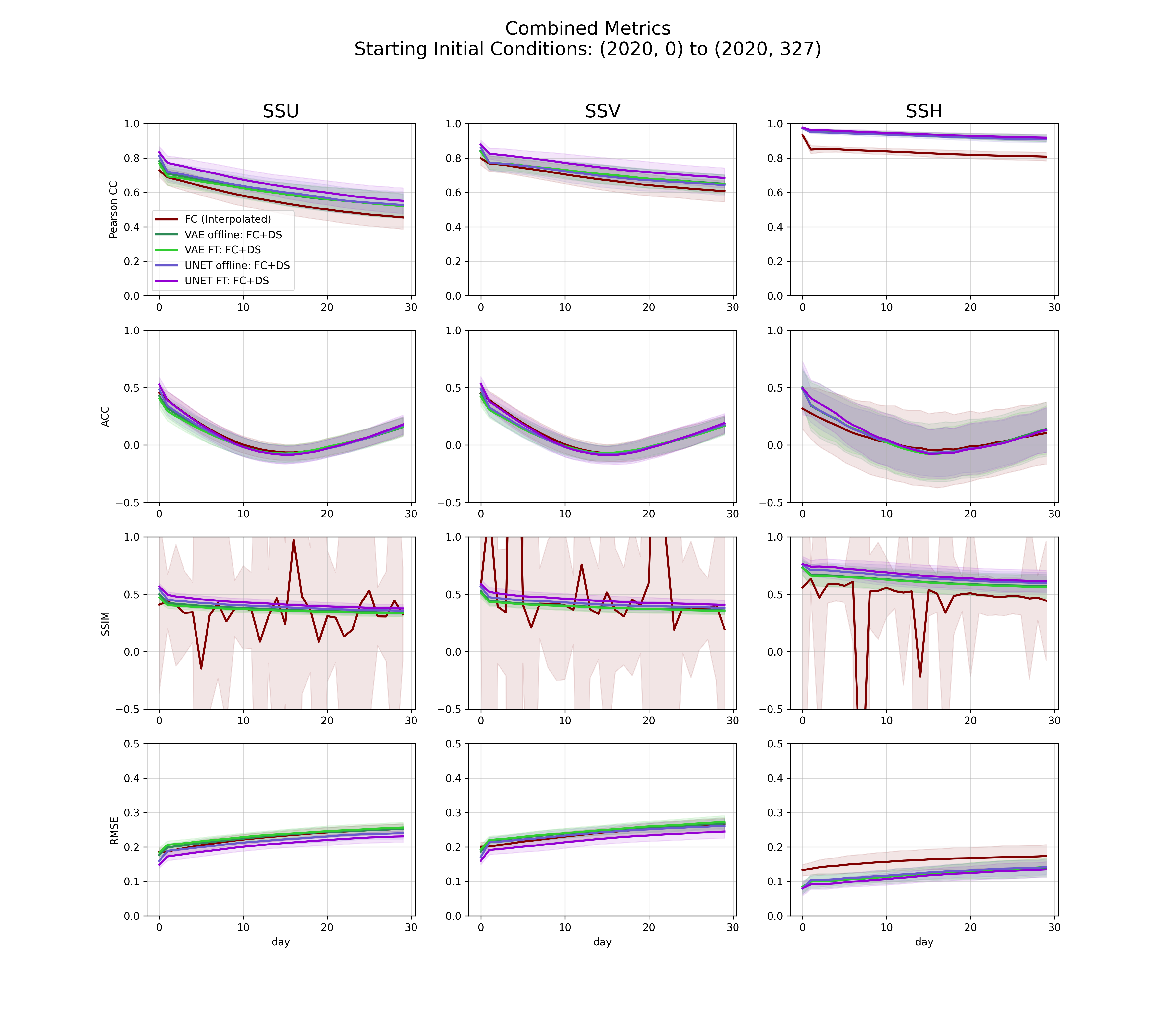}
    \caption{We compare the performance of the FCDS framework starting from $50$ initial conditions in the test dataset. First, an FNO2D model, trained with a spectral loss of \(\lambda = 0.2\), is initialized. Next, different downscaling models—either trained purely offline or offline with online fine-tuning—are applied to each FNO2D forecast (labeled as "offline" and "FT" in the legend, respectively). Each downscaling model consistently outperforms linear interpolation across the following metrics: Pearson correlation coefficient, anomaly correlation coefficient (ACC), structural similarity index (SSIM), and root mean squared error (RMSE). }
    \label{fig:model_metrics}
\end{figure}

In Fig.~\ref{fig:model_metrics}, we compare several short-term metrics to assess the performance of the FCDS framework over $30$ days. We compare the Person correlation coefficient, ACC, SSIM, and RMSE metrics for each of the different configurations of the FCDS framework with both online fine-tuned and offline VAE and UNET as the DS model. Figure~\ref{fig:model_metrics} shows that each of the configurations in the FCDS framework perform better than the navie interpolation baseline while the online fine-tuned UNET performs the best in terms of correlations and SSIM. The SSIM metric which captures the structures of the eddies and their shedding shows the instability in the interpolation-based DS model which has a very large uncertainty across the different initial conditions. In general, the FCDS framework is robust with any of the DS models and show comparable performance. It must be noted that unlike prediction tasks without downscaling the ACC does not start from $1.0$ even at the initial condition. This is because the initial conditions are from the low-resolution GLORYS dataset while the true high-resolution snapshot are from CNAPS. This is a realistic forecasting and downscaling setup where the training data and the initial condition may come from different reanalysis or observation products. 

\subsection{Power spectrum of the FCDS framework}
\begin{figure}[H]
    \centering
    \includegraphics[width=1\linewidth]{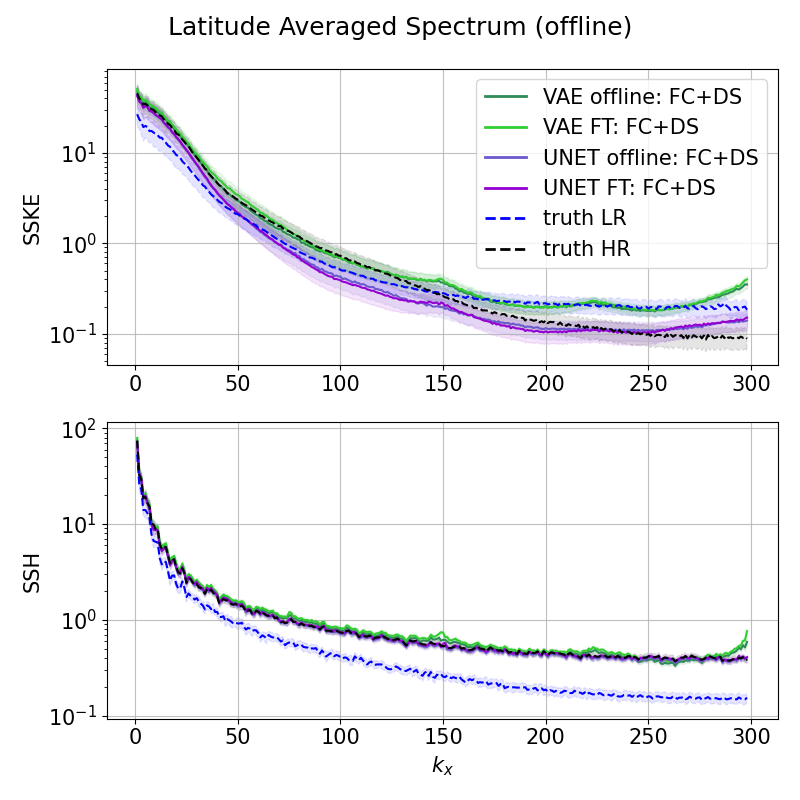}
    \caption{Offline mean power spectrum for SSKE and SSH, with standard deviation bounds shown across the ensemble of initial conditions for 30 days. The VAE captures the expected zonal SSKE spectrum more accurately compared to the UNET, however both perform similarly with respect to SSH.}
    \label{fig:model_spectrums_offline}
\end{figure}

\begin{figure}[H]
    \centering
    \includegraphics[width=1\linewidth]{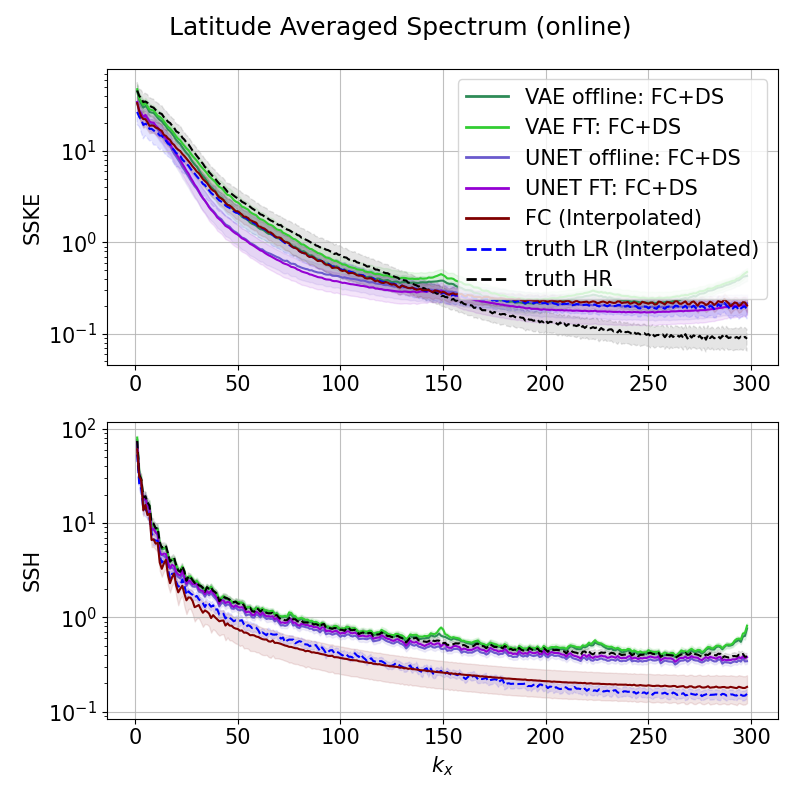}
    \caption{Online finetuned trained mean spectrum for SSKE and SSH, with standard deviation bounds shown  across the ensemble of $50$ initial conditions for $30$ days.}
    \label{fig:model_spectrums_online}
\end{figure}

In this section, we go beyond the traditional statistical short-term metrics and investigate physics-informed metrics such the kinetic energy's power spectrum and the power spectrum of the SSH fields and compare it with the high-resolution CNAPS's power spectrum. In general, for the offline FCDS framework (i.e., before executing prediction and downscaling from an initial condition in the test set) in Fig.~\ref{fig:model_spectrums_offline}, we see that the generative VAE model captures the SSKE power spectrum more accurately than the deterministic UNET-based DS model, although in either case, we see that the higher wavenumbers are not accurately captured. We further see that each of the FCDS configuration captures the SSH spectrum accurately over 30 days of prediction. However, in the online mode as shown in Fig.~\ref{fig:model_spectrums_online}, the advantage of the VAE model over the UNET model for DS disappears and they  perform similarly for both SSKE and SSH. This difference between offline and online performance of deep learning models is not new and has been reported in studies involving subgrid-scale models and parameterizations in climate models as well~\cite{lin2023systematic}. It must be noted that in online mode, the FCDS framework outperforms naive interpolation in terms of the power spectrum ensuring that the DS model does indeed capture physically realistic high-wavenumber features in the flow during autoregressive prediction and downscaling. 

\subsection{Long-term stability, mean, and variability}
\begin{figure}[H]
    \centering
    \includegraphics[width=1\linewidth]{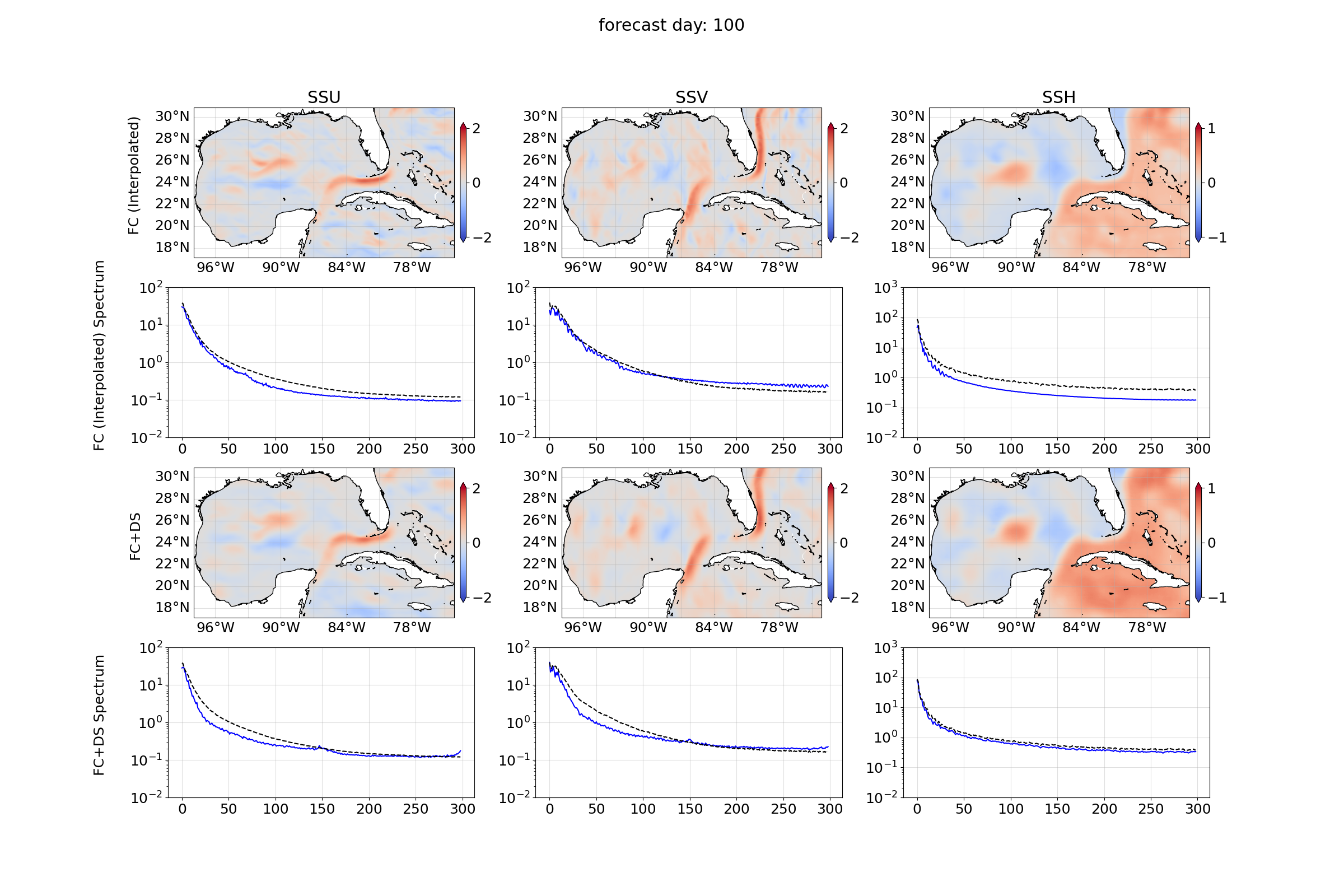}
    \caption{FCDS emulation with the online-finetuned UNET-based DS model at $100$ days. Rows from top to bottom: FC with interpolation and spectrum, FCDS and spectrum. The dotted black line is the true high-resolution spectrum, while the blue line is the spectrum from the FC with interpolation and the UNET DS model.}
    \label{fig:fcds_100day}
\end{figure}

\begin{figure}[H]
    \centering
    \includegraphics[width=1\linewidth]{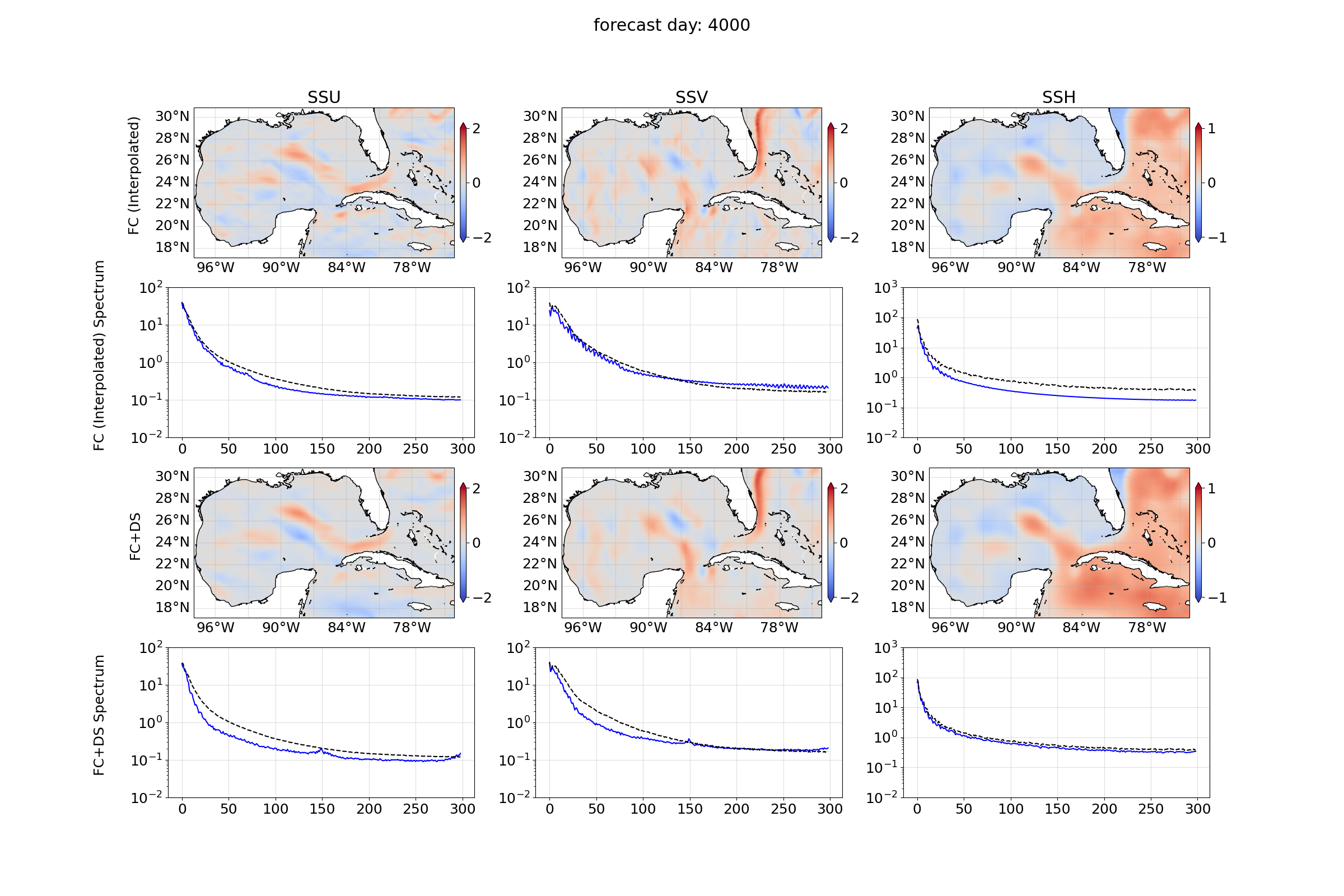}
    \caption{Same as Figure~\ref{fig:fcds_100day} but for FCDS emulation at day $4000$.}
    \label{fig:fcds_4000day}
\end{figure}

\begin{figure}[H]
    \centering
    \includegraphics[width=1\linewidth]{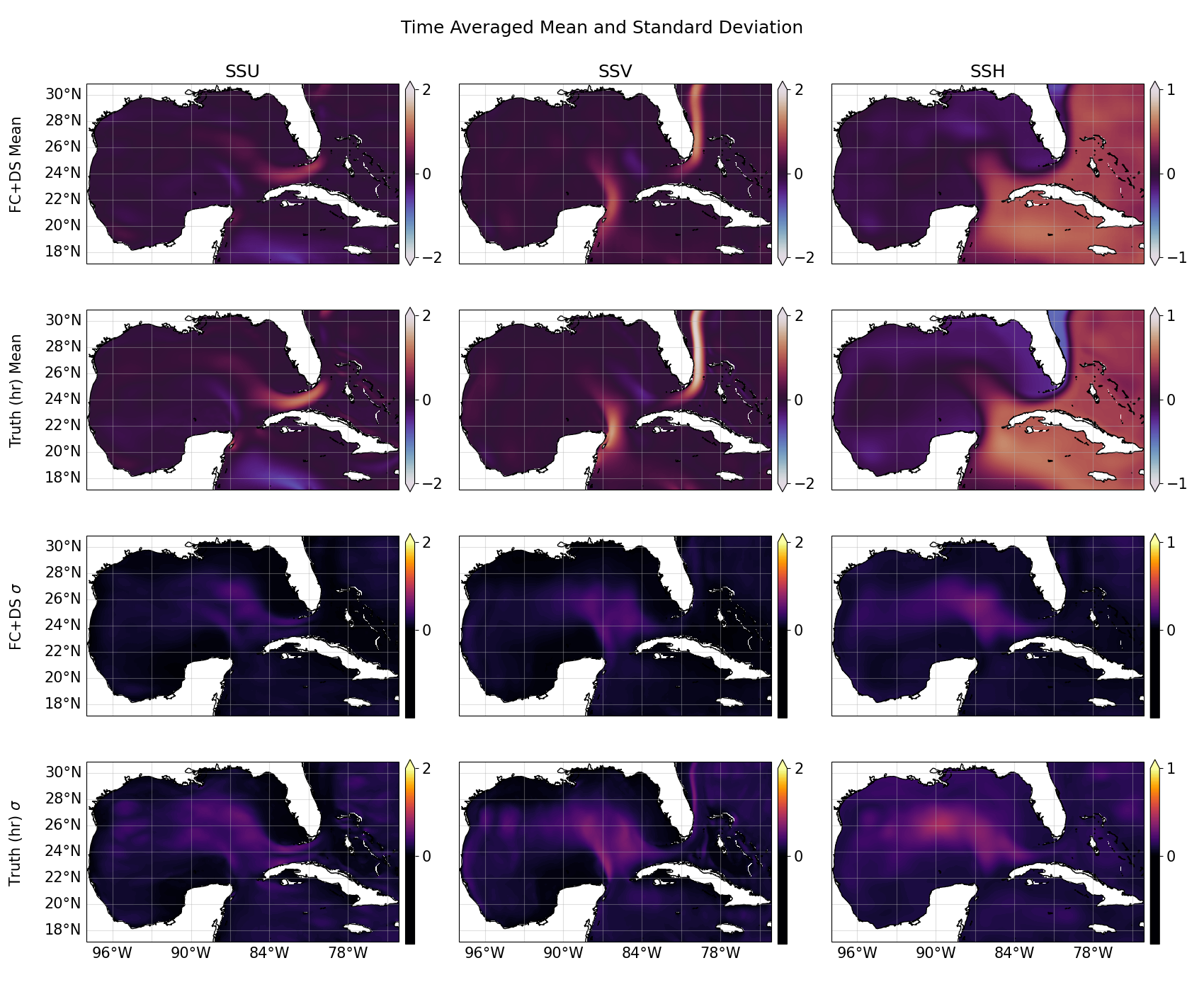}
    \caption{$10$-year mean and standard deviation of fields for FCDS prediction and truth. The first two rows are the $10$-year mean of FCDS and CNAPS (truth) and the last two rows represent the standard deviation ($\sigma$).}
    \label{fig:long_term_mean}
\end{figure}

In this section, we run the FCDS framework for $10$ years and analyze the emulations to inspect instability and unphysical drifts in the model. Figures~\ref{fig:fcds_100day} and \ref{fig:fcds_4000day} show the outputs from the FCDS framework after $100$ and $4000$ days of emulation. As can be seen, the power spectrum of SSU ans SSV remains roughly similar to the high-resolution CNAPS (with obvious artifcats near the higher wavenumbers) while the SSH spectrum is accurately captured. Finally, we  compute the long-term mean and standard deviation of the FCDS framework in Fig.~\ref{fig:long_term_mean} and compare with the mean and standard deviation of high-resolution CNAPS. Fig.~\ref{fig:long_term_mean} shows that both mean and standard deviation is accurately captured in the FCDS framework ensuring that we have a long-term physical ocean climate without drifts.

\section{Discussion}
In this paper, we introduce FCDS -- a framework for autoregressive emulation and simultaneous downscaling and bias correction for the region of GoM. We have considered only the ocean surface variables in this paper, noting that the framework is easily extendable to other sub-surface variables. The framework is purely data-driven and hence can be executed at orders of magnitude faster runtime than physics-based modeling and downscaling frameworks. Both the autoregressive model and the downscaling models are equipped with a physics-inspired spectral loss function that remedies the adverse effects of spectral bias which leads to instability and unphysical drifts~\cite{chattopadhyay2023long,bonavita2023limitations}. 

In this work, intead of taking physics-based forecasts and downscaling them to high-resolution we develop an autoregressive emulator which is orders of magnitude faster than a physics-based model and downscale the emulated fields into higher resolution. As such, the autoregressive model deviates from the true low-resolution fields due to compounding model error and chaos. Hence, the downscaling model is fine-tuned to perform bias correction as well. The bias correction strategy accounts for three sources of error: the deviation of the emulator, the discrepnacy in the resolutions between the GLORYS and CNAPS data, as well as the underlying differences in the physical parameterizations between the CNAPS and GLORYS's numerical models. 

In this work, we emulate low-resolution fields instead of the high-resolution fields. This choice stems from how spectral bias -- a fundamental bias in deep neural networks to resolve high-wavenumber features of the fields' evolve during autoregressive integration. Generally, for weather and climate emulators that have gained popularity in the climate sciences, high-resolution emulators, evolve at much lower resolution effectively, due to spectral bias ~\cite{bonavita2023limitations, chattopadhyay2023long}. This is a major cause of unphysical drifts and instabilities in these models. However, spectral bias is usually much lower when the field itself is low resolution and is thus less susceptible to instability or drifts. Hence, we focus on low-resolutiin autoregressive emulation and use a downscaling model to bring the predicted fields to higher resolution. Since, both the models are data-driven, they are very cheap to execute in inference mode; thus the final high-resolution outputs from the FCDS framework takes much less computational time than a physical model. 

Unlike other super-resolution tasks, where low resolution fields from the same distribution are downscaled to high-resolution, e.g., downscaling low-resolution ERA5 data to high-resolution ERA5, our work considers downscaling from one data distribution (GLORYS) to another (CNAPS) with different underlying physical models, parameterizations, as well as data assimilation stratgeies. This is a more realsitic set up, where the emulator is initialized with surface fields from GLORYS but the final downscaled outputs are compared with a different reanalysis product, e.g., CNAPS. While a more challneging tasks, offline downscaling and online fine-tuning together provides accurate forecasts both in short term as well as correct long-term statistical metrics that are physically consistent. 

While the FCDS framework uses surface ocean variables and remains stable at decadal time scales, there are some key components missing in the autoregressive model. To begin with, the model does not consider any atmospheric forcing which can only be incorporated with a separate atmospheric emulator. We are currently working towards a regional coupled emulator at high resolution. Furthermore, while we can successfully emulate the control climate, we cannot estimate regional response of the ocean to $CO_2$ forcings. In future work, we would focus on developing the emulator with sub-surface ocean variables as well as radiative forcing to study the climate change impacts on the GoM region.  

\section*{Open Research Section}
The codes and accompanying data for this paper is available on Zenodo under DOI: 10.5281/zenodo.14607130.

\acknowledgments
AC and SH designed the study. LJ wrote the codes and conducted the analysis. MD wrote the VAE used in the study. TW, MG, and RH provided the CNAPS data. AC, SH, LJ, and MD analyzed the results and discussed the conclusions. AC, LJ, and MD wrote the manuscript. All authors edited the manuscript. AC, MD, and LJ were supported by the National Science Foundation (grant no. 2425667) and computational support from NSF ACCESS MTH240019 and NCAR CISL UCSC0008, and UCSC0009. A part of this study was conducted when LJ was an internee at Fujitsu Research of America under the supervision of SH and AW.

%
%


%
%
%
%
%


\bibliography{Main}

\end{document}